# The potential of large germanium detector arrays for solar-axion searches utilizing the axio-electric effect for detection

F.T. Avignone III

Department of Physics and Astronomy, University of South Carolina
Columbia, South Carolina 29208, USA

The sensitivities of large arrays of low-background germanium detectors for solar axion, or axion-like particle, searches are investigated in the context of several coupling scenarios. A search was made for the 14.4-keV axion branch from the M1-transition in the thermally excited $^{57}$Fe in the solar core in $80 \, kg \cdot d$ of data from the IGEX Dark Matter experiment. In one scenario, the direct couplings of axions to hadrons and to electrons were assumed to be in the range of Peccei-Quinn (PQ) scale, $3 \times 10^5 \leq f_{PQ} \leq 7 \times 10^5 \, GeV$. This scenario is excluded by the IGEX data. In a second case, at the same PQ scale, coupling to photons and to electrons are suppressed. The IGEX data partially excludes this scenario depending on the degree of suppression. In the third case, the PQ-scale for all couplings is assumed to be in the range: $3 \times 10^6 \leq f_{PQ} \leq 7 \times 10^6 \, GeV$. Detection rates in Ge for all three scenarios are presented for several detector arrays.

**INTRODUCTION**

It is well known that strong force as described in QCD has a term that violates CP symmetry. The strength of this term is fixed by a constant, $\Theta$, that must be of order $10^{-10}$ for QCD to be consistent with the current experimental limit on the neutron electric moment, $|d_n| \leq 2.9 \times 10^{-26} e \cdot cm$ [1]. The fact that the strong interaction parameter, $\Theta$, would have such a small value is indeed unnatural. To ameliorate this situation, Peccei and Quinn postulated a new global U(1) symmetry which is spontaneously broken at a high energy scale [2]. The result of this symmetry breaking is a Goldstone boson, the axion [3,4]. For about 30 years experimentalists have been searching for this particle to confirm this elegant solution of the strong-CP problem. See for example [5] for many references to that long history. The best experimental bound on axion and axion-like particle (ALP) coupling to photons is provided by the results of the CAST experiment [6]. The tightest bounds, however, are provided by astrophysical constraints. These are discussed in detail in reference [5]. One of the primary conclusions of reference [5] is that conventional Peccei-Quinn axions with masses above a few tens of milli-electron Volts, are essentially excluded by astrophysical data (See figure 2 of reference [5]). According to this figure, in the non-astrophysical bounds, or strictly laboratory bounds, there are gaps in the constraints between 30 eV and 1 keV, and between 1 and 20 eV, i.e., between the Telescope and CAST bounds. We also explore the small window in the SN 1987A bounds, which however was recently closed by astrophysical data [5].

Experiments are motivated by the fact that theoretically axions can couple directly to hadrons, to electrons, or to photons. One of the main focuses here will be the less frequently considered direct coupling to hadrons and to electrons for the purpose of designing experimental searches. This choice motivates experimental methods that differ from the most common approaches, namely coupling to photons via the Primakoff diagram.

We examine two ranges suggested by the diagram shown in Figure 2 of reference [5], a lower mass scale, namely $(3 \times 10^5 \, GeV \leq f_{PQ} \leq 7 \times 10^5 \, GeV)$, left open in the range constrained by the SN-1987A data due to axion trapping in the supernova core, and a higher mass scale, namely $\{3 \times 10^6 \leq f_{PQ} \leq 7 \times 10^6 GeV\}$, between the CAST exclusion and axion mass interval excluded by telescope observations [5]. Here, $f_{PQ}$ is the Peccei-Quinn scale in GeV. In what follows, we test three hypotheses. In some of these scenarios the searches actually occur outside of the conventional PQ-axion model space, and the subjects of these searches are axion-like particles or ALPs.

**First**: We assume that the PQ-scale is that corresponding to the lower mass scale (stronger coupling), and that CAST has not detected axions via the axion-to-photon conversion by the Primakoff process in their magnet, because of the loss of coherence in the magnet. Even when filled by gas coherence is lost for $m_a \approx 1eV$, whereas the open window at the low-mass PQ-scale covers the range $\{10eV \leq m_a \leq 20eV\}$. Perhaps axions in this range have not been seen in other experiments searching for solar axions generated and/or detected by the Primakoff process because coupling to photons is suppressed, possibly by mechanisms suggested independently by Kaplan [7] and by Srednicki [8]. In this case, we assume that the interaction of axions that couple directly to both electrons and hadrons is not suppressed. Of course, this assumption includes axion masses outside the range of those of the conventional PQ-axion models.

**Second**: Again we assume that the PQ-scale is that corresponding to the lower mass scale, but in this case it is assumed that the direct coupling to electrons is suppressed while that of the direct coupling to hadrons is not. This affects the detection rate but not the generation rate in the sun in the 14.4-keV M1 transition in excited $^{57}$Fe in the solar core.

**Third**: In this case, we assume that both couplings to electrons and to hadrons correspond to the higher-mass PQ-scale, and hence are much weaker. The reason that CAST has not been able to cover this window is again because the corresponding axion mass range, $\{1.0eV \leq m_a \leq 2.0eV\}$, is well above that for which the CAST magnet can maintain coherence. It should be noted, however, that this range has recently been excluded by the constraint on the maximum hot dark matter population as in the case of massive neutrinos. See the discussion and Figure 3 in reference [5].

If the first and second hypotheses would agree with nature, experimental discovery of these illusory Goldstone-Bosons should be straight forward with large arrays of Ge detectors. In the case of the third hypothesis, discovery might be far more challenging, depending on the actual value of $f_{PQ}$. The scenarios defined by the above assumptions are examples that should motivate attempts to detect the axions generated in the sun from the $14.4keV$ M1 nuclear transition in $^{57}Fe$, in the solar core, with a large detector array that utilizes the axioelectric effect [9]. Corrections to the axio-electric cross sections in refs. [9,10], were recently made by Pospelov and his co-workers [11]. In the relativistic regime, the focus of this work, these corrections increased the cross sections by a factor of 2. However, in the case of the application to the non-relativistic regime, as applied by Bernabei et al., [12] these corrections were very important. Several experimental articles utilizing the axio-electric effect have already appeared in the literature [10,12-14]. Reference [10] describes the first small pilot experiment, and was small in volume and higher in background than presently achievable. Reference [12] describes a search for the absorption of cold dark-matter axions in the galactic halo with the large DAMA/LIBRA NaI array. However, in reading reference [12], it is necessary to take into account the corrections reported in reference [11]. Reference [13] was a search for the $478 keV$ axions from that M1 transition in thermally excited $^7Li^*$ in the solar core, with the Borexino detector. Reference [14] describes another small pilot experiment with a Ge detector that searched for the monochromatic axion from $14.4 keV$ M1 transition in $^{57}Fe$ in the solar core. There is a long list of experimental attempts to discover axions. At this time, however, the best laboratory bound on any of the axion-like particle (ALP) couplings has been provided by the CAST experiment, which set the upper bound, $g_{a\gamma\gamma} \leq 1.16 \times 10^{-10} GeV^{-1}$ for ALPs with masses $m_a \leq 10^{-2} eV$ [15]. It should be recognized, however, that the strongest bound was set using astrophysical data [5,16]. The CAST scenario involves the creation of axions in the sun, as well as the conversion to photons in a magnetic field, both via the Primakoff diagram. If this process is indeed suppressed [7,8], this could explain the null result for low-mass axions, while still allowing the possibility of discovery with techniques not involving this mechanism. As stated above, magnetic helioscopes suffer from the loss of coherence between the axion and photon components of their wave function superposition in the magnetic field, when the axion mass is $10^{-2} eV$ or more. Accordingly, it will be very difficult for magnetic helioscopes to reach very far into the conventional PQ-axion-model space before losing coherence.

## AXIONS FROM THE 14.4-keV M1 TRANSITION IN $^{57}$Fe IN THE SUN

The formalism for axion production, as a branch competing with M1 electromagnetic transitions, was developed by Haxton and Lee [17] to investigate a possible mechanism for cooling stars, and was also applied to a laboratory search for axions from an M1 transition in strong radioactive source of $^{65}$Zn [18]. In this experiment, the detection utilized axion-to-photon conversion, via the Primakoff diagram in a low background Ge detector. Its sensitivity was at the PQ-scale of about 250 GeV, which has long since been excluded.

Based on the work of Haxton and Lee, Moriyama developed the concept of searching for these axions by exciting a target of $^{57}$Fe in a detector on earth [19]. A very small pilot experiment was done by another group [20], by placing a small foil of iron enriched in $^{57}Fe$ on a Ge detector and searching for the gamma ray from the decay of the $14.4 keV$ gamma ray from the state excited by the axion. However, no suitable detector has been developed, with a large enough mass to make a meaningful search. It would be very desirable to exploit this technique, because it depends on the direct coupling of axions to nucleons for both production and detection, avoiding dependence on other couplings.

The Lagrangian describing the coupling of axions to nucleons is given in reference [17]:

$$L = a\overline{\Psi}i\gamma_5(g_0\beta - g_3)\Psi, \qquad (1)$$

where $g_0$ and $g_3$ are the isoscalar and isovector coupling constants respectively. The axion to photon branching ratio for $M1$ transitions was derived by Haxton and Lee [17], and is written as follows:

$$\frac{\Gamma_a}{\Gamma_\gamma} = \frac{1}{2\pi\alpha(1-\delta^2)} \left[ \frac{g_0 \beta - g_3}{(\mu_0 - 1/2)\beta + \mu_3 - \eta} \right]^2. \quad (2)$$

In (2), $\mu_0$ is the isoscalar magnetic moment $(\mu_0 - 1/2) \approx 0.38$, while $\mu_3 \approx 4.71$, is the isovector magnetic moment. The parameters, $\eta$ and $\beta$, are nuclear structure dependent quantities calculated in reference [17]. The parameter $\beta \approx +1$ for an unpaired proton, and $\beta \approx -1$ for an unpaired neutron, while $\eta \approx +0.80$ in the case of $^{57}Fe$, which has an unpaired neutron, and for example, $\eta \approx -3.74$ for $^{55}Mn$ and $-1.20$ for $^{23}Na$, both of which have unpaired protons [17].

The axion-nucleon coupling constants depend on several parameters as given in equation (3) below [17,18]. The isoscalar and isovector couping constants are:

$$g_0 = -1.61 \times 10^{-7} \left\{ \frac{3F - D + 2S}{f_{PQ}} \right\}, \quad \text{and} \quad g_3 = -4.84 \times 10^{-7} \left\{ \frac{(D+F)(1-z)}{f_{PQ}(1+z)} \right\}. \quad (3)$$

In equation (3), $F$ and $D$ are invariant matrix elements of the axial current with values: $F \approx 0.48$ and $D = 0.77$ [17]. Also, $z = m_u/m_d \approx 0.56$ in the quark model, and the Peccei-Quinn mass scale, $f_{PQ}$, in this case, is given in units of $10^6 GeV$. The quantity, $S$, is the flavor-singlet axial-vector matrix element, and plays a crucially important role in calculating both solar-axion fluxes and interaction cross sections.

**COMPUTATION OF THE AXION FLUX ON EARTH FROM THE 14.4 KEV $M1$ TRANSITON IN $^{57}$Fe IN THE CORE OF THE SUN**

The axion flux can be calculated based on reference [19]. The resulting expression is:

$$\frac{d\Phi_a}{dE_a} = 2.0 \times 10^{13} cm^{-1} s^{-1} keV^{-1} \left( \frac{10^6 GeV}{f_{PQ}} \right)^2 C^2 \quad (4)$$

where,

$$C = -1.19 \left( \frac{3F - D + 2S}{3} \right) + (D+F)\left( \frac{1-z}{1+z} \right). \quad (5)$$

In our analyses, we use $\{0.35 \leq S \leq 0.55\}$, which includes the entire overlap of the ranges of experimental values $(0.37 \leq S \leq 0.53)$, (95%CL) from Altarelli et al.,

[21], and that of the Spin-Muon Collaboration, $(0.15 \leq S \leq 0.50)$ [22]; accordingly $\{0.037 \leq C^2 \leq 0.125\}$. The fluxes of the 14.4-keV axions can be determined from equation (4) by realizing that the Doppler broadening in the solar core would result in an axion line with a width of 5-eV. Accordingly:

$$\Phi_a(14.4 keV) = 1.0 \times 10^{11} cm^{-2} s^{-1} \left(\frac{10^6 GeV}{f_{PQ}(GeV)}\right)^2 \{C(F,D,S)\}^2 \qquad (6)$$

Equation (6) was used to calculate the fluxes for the following range of parameters: $\{3 \times 10^5 GeV \leq f_{PQ} \leq 7 \times 10^5 GeV\}$, and $\{0.35 \leq C(F,D,S) \leq 0.55\}$, which includes the entire region of overlap of the two ranges in the parameter, S, reported in references [21,22]. The values of the flux are tabulated in Table 1. To obtain the predicted values of $\Phi_a(14.4 keV)$ for the axion couplings between that of the CAST bounds, and that by the telescope experiments, namely, $\{3 \times 10^6 GeV \leq f_{PQ} \leq 7 \times 10^6 GeV\}$, simply multiply the values in Table 1 by $10^{-2}$.

**THE AXIO-ELECTRIC EFFECT AS THE PROPOSED DETECTION MECHANISM**

The formalism for the axio-electric effect was given by Dimopoulos et al., [9], and later applied to a pilot experiment to search for axions generated by the Primakoff process in the solar core [10]. The relevant expressions relating the axio-electric and photo-electric effect, corrected according to reference [11], are:

$$\sigma_{ae} = \frac{\alpha_{axion}}{2\alpha_{EM}} \left(\frac{\hbar\omega}{m_e c^2}\right)^2 \sigma_{photo-electric} \quad \text{and} \quad \alpha_{axion} = \frac{1}{4\pi}\left(\frac{2x'_e m_e c^2}{f_{PQ}}\right)^2 \qquad (7)$$

$$\text{where } x'_e \approx 1, \text{ and } \alpha_{axion} = \frac{8.312 \times 10^{-8} GeV^2}{f_{PQ}^2} \qquad (8)$$

$$\sigma_{axio-electric} = \frac{2.18 \times 10^{-11} GeV^2 \times (E_a^2)}{f_{PQ}^2} \times \sigma_{photo-electric} \qquad (9)$$

In equation (9), $E_a$ is in keV and $f_{PQ}$ is the Peccei-Quinn scale in GeV. It should be noted that the parameter, $f_{PQ}$, used in this work is defined such that $f_{PQ} = 6 \times 10^6 GeV$ corresponds to an axion mass of 1 eV. However, in this paper the same parameter is used in scenarios outside of the model space of the standard PQ-axion. The early search described in reference [10], was only a small pilot experiment, and the application of this technique to a large array of low-background Ge detectors would be far more effective, particularly since the $14.4 keV$ axion line would give a sharp peak when the axioelectric effect is utilized.

The photo-electric cross section for 14.4-keV photons on Ge from standard NIST tables is: $\sigma_{photo} = 1.2 \times 10^{-20} cm^2$. Applying equation (8), the axio-electric cross sections were computed and used, with the values for the flux given in Table 1 to compute the detection rates in Table 2. To obtain the rates for the **Second** hypothesis multiply the rates in the cross section and rate tables by $10^{-2}$. To obtain the predicted rates for the **Third** hypothesis, multiply the fluxes in Table 1 by $10^{-2}$ and the rates in Table 2 by a factor of $10^{-4}$.

**WHAT CAN THE RAW DATA FROM THE IGEX DARK MATTER EXPERIMENT TELL US ABOUT SOLAR AXIONS?**

In 2002 the IGEX collaboration published their results from a search for Weakly Interacting Massive Particles (WIMPS) using low-energy data from a 2-kg ultralow background Ge $\beta\beta$–decay detector operated in the Canfranc Underground Laboratory in Canfranc, Spain [23]. This body of data represented 80 $kg \cdot days$ of exposure. Between 10 and 20 keV, the average background was ~0.04 $counts \cdot keV^{-1} kg^{-1} d^{-1}$. The experimental details are given in reference [23]. The data in the narrow region of interest are given in Table-3 below. These data were published, however, were not previously used to search for axions.

In the case of the **First** hypothesis, the range of the Peccei-Quinn scales for both axion production in the sun, and detection by the axio-electric effect is, $\{3 \times 10^5 GeV \leq f_{PQ} \leq 7 \times 10^5 GeV\}$, the window not closed by the SN 1987A data because of axion trapping in supernova core by strong coupling to nucleons. In this scenario, there should be between 0.58 and 58.4 $counts \cdot keV^{-1} kg^{-1} d^{-1}$, or a minimum of 46-counts in $80 kg \cdot d$, in a peak at $14.4 keV$. A very simple Poisson-fluctuation analysis implies that there are $\leq 7.4$ counts (99.73% CL). This window would then be completely closed by these data, but only in the case that the Peccei-Quinn scale for the direct coupling of axions to electrons is not suppressed by mechanisms not included here. A larger detector array could probe that question with the other scenarios described below. It should be noted that that window was recently closed by limits o hot dark matter [5].

In the case of the **Second** hypothesis, we assume that the direct coupling of axions to electrons is suppressed to effectively place the range of PQ-scale, $f_{pq}$, in the heavier mass range, such that the flux remains the same as that given in Table-1, but the cross sections, and the resulting detection rates given in Table 2 are reduced by factors of $10^{-2}$. This would result in detection rates in a $60 kg$ array of Ge detectors, for example, of between about 127 and 12,700 detected events per year. In the $80 kg \cdot d$ IGEX data these rates would be about $\{0.5 - 47\}$, much of which is already excluded by the data. Suppose, instead, the coupling to electrons is suppressed to a PQ-scale in the range of the CAST exclusion band, namely $\{6 \times 10^6 \leq f_{PQ} \leq 2 \times 10^7\} GeV$. Then the rates in the $60 kg$ array would be between about 15 and 800 detections per year. In this case, there would be between about 0.0 and 3 counts in the $80 kg \cdot d$ of existing IGEX data.

If we consider the **Third** hypothesis, namely that the window is the one that lies between the exclusion region of the CAST experiment, and that of the telescope data namely $\{3 \times 10^6 GeV \leq f_{PQ} \leq 7 \times 10^6 GeV\}$, we predict event rates between about $6.0 \times 10^{-5}$ and $6.0 \times 10^{-3}$ $kg^{-1}d^{-1}$, respectively, which would require a large detector array to observe any counts due to axion interactions. For example, if a 1000 kg Ge detector array would operate for one year, the values in Tables 2 and 3 would predict the detection of between 21 and 2130 events per year, depending on the values in the range nature selects for $f_{PQ}$ and $S$. For a $60 kg$ array, this range of rates is from about 1 to 130 per year. The data from a 60 kg array would be very interesting, and if no signal were seen, longer operating times and larger arrays would be appropriate.

**CONCLUSIONS**

It is clear from the above discussions that large low-background arrays of Ge detectors would allow a significant improvement in the sensitivities for searches for hadronic axions that interact directly with nuclei, and also directly with electrons via the axio-electric effect. In addition, the signal would experience a model-independent annual modulation, with a maximum in January and minimum in July, because the sun to earth distance factor $(4\pi r^2)^{-1}$ changes about 6.9% between January and July. The efficacy of such an array for this purpose will depend on the energy resolution and threshold, as well as on the background in the region around 14.4-keV. The new point-contact detectors, like those used in the very recent paper by Aalseth et al., [24], were able to observe the Ga L-x-ray at 1.29-keV. The energy resolution at 10-keV was about 117-eV. Just this dramatic improvement in energy resolution, over the usual 1-keV resolution of convential detectors, would improve the sensitivity as much as lowering the background by about one order of magnitude. Both the MAJORANA [25] and GERDA [26] collaborations are planning large arrays of ultra-low background Ge detectors for searches for $0\nu\beta\beta$-decay. They should be encouraged to attempt to achieve backgrounds in the region of the $14.4 keV$ solar-axion line to probe the **Third** hypothesis as sensitively as possible. This would make a major improvement in sensitivity.

**ACKNOWLEDGEMENTS**

The author is grateful to Georg Raffelt for pointing out the very recent work by Pospelov, Ritz and Voloshin that required a change in the axio-electric cross section. This work was supported by NSF Grant PHY-0500337.

Table 1. Calculated fluxes for 14.4-keV solar-axions for the hadronic-axion window $\{3\times10^5 GeV \leq f_{PQ} \leq 7\times10^5 GeV\}$, and for the range of values of the flavor-singlet, axial-vector matrix element, $\{0.35 \leq S \leq 0.55\}$. The fluxes are given in units of $10^{11}$ axions cm$^{-2}$s$^{-1}$.

| $S / f_{PQ}(GeV)$ | $3.0\times10^5$ | $3.5\times10^5$ | $4.0\times10^5$ | $4.5\times10^5$ | $5.0\times10^5$ | $6.0\times10^5$ | $7.0\times10^5$ |
|---|---|---|---|---|---|---|---|
| 0.35 | 0.400 | 0.294 | 0.225 | 0.178 | 0.144 | 0.100 | 0.074 |
| 0.40 | 0.589 | 0.433 | 0.331 | 0.262 | 0.212 | 0.147 | 0.108 |
| 0.45 | 0.811 | 0.596 | 0.456 | 0.361 | 0.292 | 0.203 | 0.149 |
| 0.50 | 1.07 | 0.784 | 0.600 | 0.474 | 0.384 | 0.267 | 0.196 |
| 0.55 | 1.36 | 0.996 | 0.763 | 0.603 | 0.488 | 0.339 | 0.249 |

* To obtain fluxes for the axion window between the CAST and the telescope data, namely $\{3\times10^6 GeV \leq f_{PQ} \leq 7\times10^6 GeV\}$, multiply the values in this table by $10^{-2}$.

Table 2. Calculated detection rates for 14.4-keV solar-axions in germanium for the hadronic-axion window $\{3\times10^5 GeV \leq f_{PQ} \leq 7\times10^5 GeV\}$, and for the range of values of the flavor-singlet, axial-vector matrix element, $\{0.35 \leq S \leq 0.55\}$. The rates are given on counts/kilogram/day $(c \cdot kg^{-1} \cdot d^{-1})^*$

| $S / f_{PQ}(GeV)$ | $3.0\times10^5$ | $3.5\times10^5$ | $4.0\times10^5$ | $4.5\times10^5$ | $5.0\times10^5$ | $6.0\times10^5$ | $7.0\times10^5$ |
|---|---|---|---|---|---|---|---|
| 0.35 | 17.2 | 9.27 | 5.44 | 3.40 | 2.23 | 1.08 | 0.58 |
| 0.40 | 25.3 | 13.7 | 8.01 | 5.00 | 3.28 | 1.58 | 0.85 |
| 0.45 | 34.8 | 18.8 | 11.0 | 6.89 | 4.42 | 2.18 | 1.18 |
| 0.50 | 45.9 | 24.7 | 14.5 | 9.05 | 5.95 | 2.87 | 1.55 |
| 0.55 | 58.4 | 31.4 | 18.5 | 10.8 | 7.57 | 3.65 | 1.97 |

* To obtain the rates for a 1000-kg detector, and for the higher PQ-scale axion window between CAST and the telescope data, $\{3\times10^6 GeV \leq f_{PQ} \leq 7\times10^6 GeV\}$, multiply the values on this table by $10^{-1}$.

Table 3. Low energy data from the IGEX RG-II detector with $80 kg \cdot days$ of exposure [21].

| E(keV) | counts | E(keV) | counts |
|---|---|---|---|
| 10.5 | 12 | 15.5 | 6 |
| 11.5 | 17 | 16.5 | 8 |
| 12.5 | 12 | 17.5 | 6 |
| 13.5 | 7 | 18.5 | 1 |
| 14.5 | 6 | 19.5 | 4 |

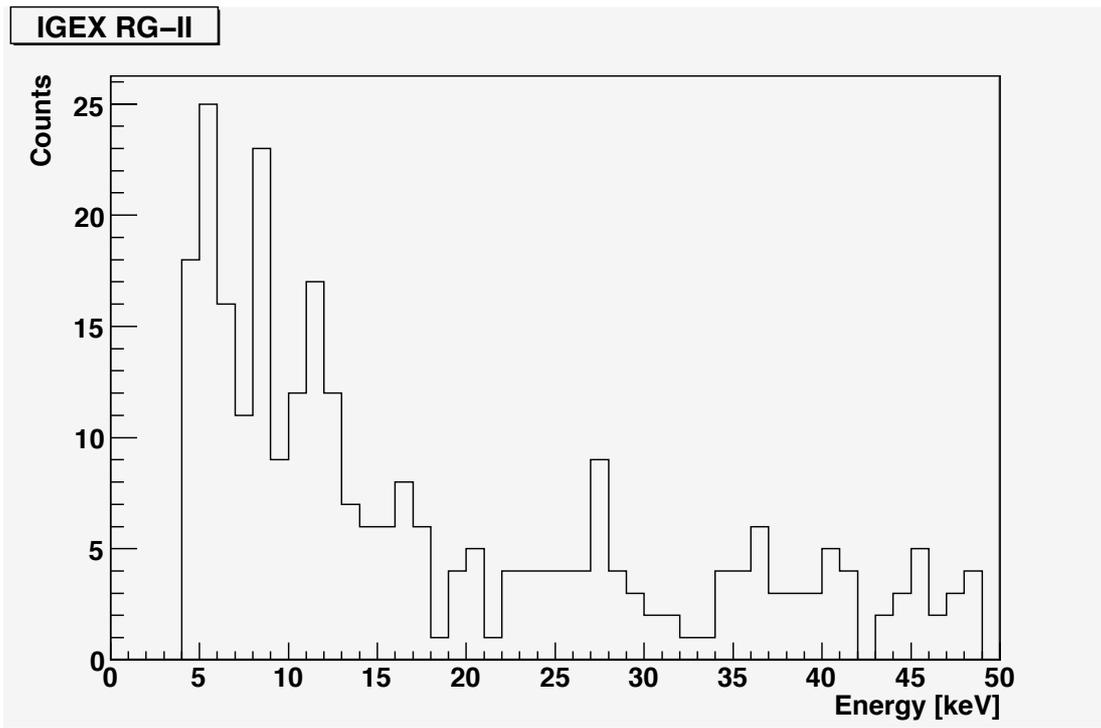

Figure 1. Histogram of $80 kg \cdot d$ of IGEX-II data from reference [21]. The numerical data from $(10-20) keV$ is given in Table 3.